\documentclass[twocolumn,prl,showpacs,preprintnumbers]{revtex4}
\usepackage{graphicx}

\newcommand{\bra}[1]{\left\langle#1\right|}
\newcommand{\ket}[1]{\left|#1\right\rangle}
\newcommand{\ov}[1]{\left\langle#1\right\rangle}
\begin{document}

\preprint{Draft version 1}

%\title{Mesoscopic System Mediated Entangling of Mesoscopic Systems}
\title{Entangling Pairs of Nano-Cantilevers,
Cooper-Pair Boxes and Mesoscopic Teleportation}
\author{S. Bose $^\ddag$ and G. S. Agarwal $^\S$}
\affiliation{$^\ddag$ Department of Physics and Astronomy,
University College London, Gower St., London WC1E 6BT, UK}
 \affiliation{$^\S$ Dept. of Physics, Oklahoma State
University, Stillwater, Oklahoma 74078-3072}

\date{26 July 2005}

%______________________________________________________________________ ABSTRACT

\begin{abstract}
We propose two schemes to establish entanglement between two
mesoscopic quantum systems through a third mesoscopic quantum
system. The first scheme entangles two nano-mechanical oscillators
in a non-Gaussian entangled state through a Cooper pair box.
Entanglement detection of the nano-mechanical oscillators is
equivalent to a teleportation experiment in a mesoscopic setting.
The second scheme can entangle two Cooper pair box qubits through
a nano-mechanical oscillator in a thermal state without using
measurements in the presence of arbitrarily strong decoherence.
\end{abstract}

%\pacs{03.67.-a, 03.65.-w, 05.30.-d}

\maketitle

%______________________________________________________________________ Article

 Probing quantum superpositions and
entanglement with mesoscopic mechanical systems has recently
developed into an area of substantial interest
\cite{zeilinger,julsgaard,deb,gsa1,bose,mancini,blencowe,bouwmeester,eisert}.
The most striking experimental demonstrations are the interferometry
of mesoscopic free particles (molecules) \cite{zeilinger} and the
entangling of mesoscopic atomic ensembles \cite{julsgaard}.
Proposals for the generation of entanglement between Bose-Einstein
condensates \cite{deb} and coherence between states of mesoscopic
atomic ensembles have been made \cite{gsa1}. Some early proposals
involving harmonically bound mesoscopic systems were based on
opto-mechanical effects where schemes for observing coherent
superpositions of states of the movable mirror \cite{bose} and
entanglement between two such mirrors \cite{mancini} were proposed.
Soon, however, a canonical system of a Cooper-pair box coupled to a
mesoscopic cantilever was introduced \cite{blencowe}. It offered an
optics-free, fully nano-technological alternative, with switchable
couplings for such schemes. Accordingly, a scheme to observe
coherent superpositions between states of a mesoscopic cantilever,
as well as its entanglement with a Cooper pair box was proposed
\cite{blencowe}. Recently, interferometric proposals to probe
superpositions of states of movable mirrors have also been proposed
\cite{bouwmeester}. Very recently, a proposal to entangle two well
separated nano-electromechanical oscillators through a harmonic
chain has also been made \cite{eisert}. A host of other quantum
effects are expected to be seen in mesoscopic mechanical systems
\cite{blencowe-review,wybourne,roukes,irish,stantamore,tunneling,sun}
including quantum computation \cite{cleland}. These theoretical
proposals are fuelled by the rapid technological progress in the
fabrication of nano-mechanical systems and experiments approaching
the quantum regime \cite{schwab-roukes-review,GHz-roukes}.

  The Hamiltonian which generates entanglement between a Cooper pair
box and a cantilever in Ref.\cite{blencowe} offers many more
exciting entangling possibilities even with {\em minimal}
additions to the number of systems, such as just one extra Cooper
pair box {\em or} just one extra cantilever. In this letter we
show that with the above minimal addition, one can entangle two
mesoscopic systems of the {\em same} dimension: two discrete
variable systems (two Cooper pair boxes) or two continuous
variable systems (two nano-mechanical cantilevers). One can also
verify their entanglement with an entanglement witness or
teleportation with higher than classically achievable fidelity. An
interesting feature of the entangling of the cantilevers is that
they are placed in a {\em non-Gaussian} continuous variable
entangled state as a result of our scheme. Till date, only
Gaussian entangled states have been used in continuous variable
implementations of quantum information processing \cite{furusawa},
and the scheme we suggest might enable one to realize a
non-Gaussian entangled state. The scheme we suggest for detection
of the non-Gaussian entanglement is equivalent to possibly the
simplest realization of a quantum teleportation experiment with
entangled nano-mechanical cantilevers. Positive features of the
entangling scheme for the Cooper pair boxes are its applicability
in entangling non-neighboring (not directly interacting) boxes in
an array and its robustness to the thermal nature as well as
decoherence of the states of the mediating cantilever.  Most
importantly, our schemes seek to extend the domain of quantum
behavior by entangling two mesoscopic systems through a third
mesoscopic system.

\begin{figure}
\begin{center}
\includegraphics[width=3.5in, clip]{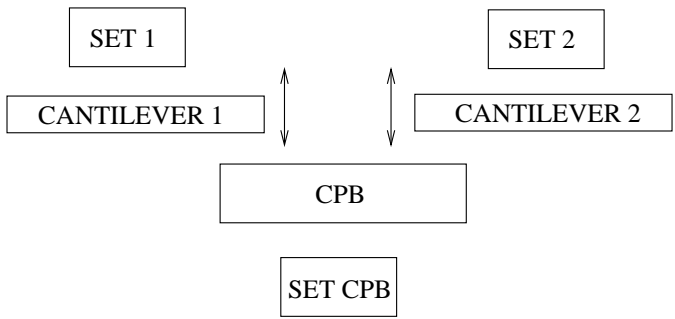}
 \caption{The figure shows a schematic diagram of the setup for entangling two cantilevers, denoted as cantilever 1 and cantilever 2
 respectively, through a Cooper pair box denoted as CPB.
 For the entangling, measurements are only needed to be performed on the CPB, which is done with the help of the single electron transistor
 SET CPB. For verification of the entanglement of cantilevers 1 and
 2 by a mesoscopic teleportation,
 measurements need to be performed on them through SET1 and SET2 respectively.}
\label{cantent}
\end{center}
\end{figure}

  {\em Entangling two nano-cantilevers:} A Cooper pair box (CPB) is an example of a qubit with states
   $\ket{0}$ and $\ket{1}$ representing $n$ or $n+1$ Cooper
   pairs in the box \cite{nakamura1,blencowe}. It can be made to
   evolve under a Hamiltonian $-\frac{E_J}{2}\sigma_x$ by the application of an appropriate voltage
   pulse \cite{nakamura1,blencowe}, where $\sigma_x$ is
   the Pauli-X operator and the parameter $E_J$ is called the Josephson
   coupling. This gives rise to coherent
   oscillations between the $\ket{0}$ and $\ket{1}$ states as
   observed in Ref.\cite{nakamura1}. A nano-mechanical
   cantilever (NC), on the other hand is a simple example of a quantum
   harmonic oscillator. We now proceed to the proposal for entangling two cantilevers
   based on their interaction with a single CPB. The setup is shown in Fig.\ref{cantent}. The
   Hamiltonian required for the scheme is given by
\begin{equation}
H=-2E_C\sigma_z+\hbar\omega_m a^{\dagger} a +\hbar\omega_m
b^{\dagger} b+\lambda \{(a+a^{\dagger})+(b+b^{\dagger})\}\sigma_z,
\end{equation}
where the parameter $E_C$ is called the charging energy of the
Cooper pair box, $\sigma_z$ is the Pauli-Z operator for the CPB,
operators $a,a^{\dagger}$ and $b,b^{\dagger}$ are the
creation/annihilation operators for two oscillators and $\lambda$
is a coupling strength. We assume that the NCs are prepared
initially in their ground state (this is quite realistic for the
GHz oscillators available now \cite{GHz-roukes} by cooling, as
suggested in Ref.\cite{cleland}). Accordingly, we start with the
cantilevers in the initial state $\ket{0}_a\ket{0}_b$, where
subscripts $a$ and $b$ denote the two cantilevers, and the CPB in
the state $\frac{1}{\sqrt{2}}(\ket{0}+\ket{1})$ (This state can be
prepared by using a voltage pulse to accomplish a $\pi/2$ rotation
about x-axis through $-\frac{E_J}{2}\sigma_x$ followed by local
phase adjustments). The evolution that takes place in a time
$T=\pi/\omega_m$ is
\begin{eqnarray}
\frac{1}{\sqrt{2}}&&(\ket{0}+\ket{1})\ket{0}_a\ket{0}_b
\rightarrow \nonumber\\
\frac{1}{\sqrt{2}}&&(e^{-i\frac{2E_CT}{\hbar}}\ket{0}\ket{-2\beta}_a\ket{-2\beta}_b\nonumber
\\&&+e^{i\frac{2E_CT}{\hbar}}\ket{1}\ket{2\beta}_a\ket{2\beta}_b),
\end{eqnarray}
where $\beta=\lambda/\hbar\omega_m$ is a dimensionless coupling
and $\ket{\pm4\beta}$ are coherent states. For simplicity, we will
assume that $\frac{2E_CT}{\hbar}$ is an integral multiple of
$2\pi$. We now measure the CPB in the basis
$\ket{\pm}=\frac{1}{\sqrt{2}}(\ket{0}\pm\ket{1})$ to obtain the
state
\begin{eqnarray}
|\psi(\pm)\rangle_{ab}&=&\frac{1}{\sqrt{2}}(\ket{-2\beta}_a\ket{-2\beta}_b
\nonumber\\ &\pm&\ket{2\beta}_a\ket{2\beta}_b), \label{state1}
\end{eqnarray}
where the upper and lower signs stand for the $\ket{+}$ and
$\ket{-}$ outcomes respectively. If $\beta \sim 1$, as will
happen, for example, if one takes the parameters of
Ref.\cite{blencowe}, then $e^{-16\beta^2}\sim O(10^{-7})$ and both
states $|\psi(\pm)\rangle_{ab}$ have nearly one ebit of
entanglement and each outcome has a probability of nearly $1/2$ to
occur. $|\psi(\pm)\rangle_{ab}$ are a class of non-Gaussian
continuous variable
  entangled states known as
  {\em entangled coherent states}, proposed originally in the optical context
  \cite{sanders}. It is trivial to check that the scheme also
  works if the cantilevers started in coherent states of non-zero amplitude.

{\em Verifying the entanglement of the cantilevers by
teleportation:} An interesting question now is how to verify the
entanglement
  of the states $|\psi(\pm)\rangle_{ab}$. The non-local character can be ascertained {\em in principle} from
  Bell's inequality experiments \cite{jeong}. However, these involve measurements in a highly non-classical (Schroedinger
  Cat-like) basis \cite{jeong},
  and could be
  rather difficult for a NC. For an NC,
  position/momentum measurements seem natural. Unfortunately, from joint uncertainties
  in position and momentum of the two NCs, the entangled nature of the state $|\psi(\pm)\rangle_{ab}$
  cannot be inferred. We will thus use quantum teleportation
  through $|\psi(\pm)\rangle_{ab}$ to demonstrate its entangled nature.
  Note that the possibility of teleportation of Schr\"{o}dinger Cat states of a third
  oscillator through the entangled coherent state of two oscillators has already been pointed
  out by van Enk and Hirota \cite{vanEnk} in the quantum optical context. However, for NCs,
  preparing a third NC in a highly non-classical state such as a Schr\"{o}dinger cat is
  challenging,
  making it directly interact with one of the entangled NCs is
  difficult and moreover, we do not want to increase the
  complexity of the system by adding an extra NC.
  We will thus concentrate on the
  teleportation of the state of a qubit through
  $|\psi(\pm)\rangle_{ab}$ with better than classically achievable ($2/3$) fidelity.
This will prove the entangled nature of the state
$|\psi(\pm)\rangle_{ab}$.

   For the teleportation protocol, first assume that the NCs were prepared in $|\psi(+)\rangle_{ab}$ as a result
   of the measurement of the CPB in the $\ket{\pm}$ basis. The CPB is now, of course, disentangled from the state
   of the NCs. It is thus now prepared in the
   arbitrary state $\cos{\theta/2} \ket{0}+e^{i\delta} \sin{\theta/2}
   \ket{1}$ which we want to teleport through $|\psi(+)\rangle_{ab}$. The CPB interacts with cantilever $a$ for a time $T$ and
   the resulting evolution is:
   \begin{eqnarray}
(\cos{\theta/2}\ket{0}&+&e^{i\delta}
\sin{\theta/2}\ket{1})|\psi(+)\rangle_{ab}
\rightarrow \nonumber\\
\frac{1}{\sqrt{2}}&&(\cos{\theta/2}\ket{0}\ket{0}_a\ket{-2\beta}_b\nonumber\\
&+&e^{i\delta}
\sin{\theta/2}\ket{1}\ket{4\beta}_a\ket{-2\beta}_b\nonumber
\\&+&\cos{\theta/2}\ket{0}\ket{-4\beta}_a\ket{2\beta}_b
\nonumber\\&+&e^{i\delta}
\sin{\theta/2}\ket{1}\ket{0}_a\ket{2\beta}_b). \label{ev1}
\end{eqnarray}
The position of the cantilever $a$ and the state of the CPB in the
$\ket{\pm}$ basis are now measured. All the above corresponds to
the Bell state measurement part of the teleportation procedure. As
$e^{-8\beta^2}<<1$, there is a probability $\sim 1/2$ that the
cantilever is projected to the state $\ket{0}_a$. Let us, for the
moment, concentrate on this outcome. Contingent on this outcome,
the state of the CPB is projected to $\ket{+}$ and $\ket{-}$ with
$1/2$ probability each, corresponding to which the state of
cantilever $b$ goes to $\cos{\theta/2}\ket{-2\beta}_b+e^{i\delta}
\sin{\theta/2}\ket{2\beta}_b$ and
$\cos{\theta/2}\ket{-2\beta}_b-e^{i\delta}
\sin{\theta/2}\ket{2\beta}_b$. Let us assume the state to be
$\cos{\theta/2}\ket{-2\beta}_b+e^{i\delta}
\sin{\theta/2}\ket{2\beta}_b$ for the moment. In some sense the
above state of cantilever $b$ already contains the teleported
quantum information from the original state of the CPB. However,
it is difficult to verify this information while it resides in the
state of cantilever $b$. So we map it back from cantilever $b$ to
the CPB (which is now disentangled as a result of the previous
measurement) by preparing the CPB in the state $\ket{+}$, allowing
for the evolution
\begin{eqnarray}
\ket{+}(\cos{\theta/2}\ket{-2\beta}_b&+&e^{i\delta}
\sin{\theta/2}\ket{2\beta}_b) \rightarrow\nonumber\\
\frac{1}{\sqrt{2}}(\cos{\theta/2}\ket{0}\ket{0}_b&+&e^{i\delta}
\sin{\theta/2}\ket{1}\ket{4\beta}_b \nonumber\\
\cos{\theta/2}\ket{0}\ket{-4\beta}_b&+&e^{i\delta}
\sin{\theta/2}\ket{1}\ket{0}_b),\label{ev2}
\end{eqnarray}
and then measuring the position of cantilever $b$. With a
probability $1/2$ it is $\ket{0}_b$, for which the CPB is
projected to the state $\cos{\theta/2} \ket{0}+e^{i\delta}
\sin{\theta/2}
   \ket{1}$, thereby concluding a chain of operations leading to
teleportation with unit fidelity. In the case when the outcome
$\ket{-}\ket{0}_a$ is obtained during the Bell measurement
procedure, a teleportation with unit fidelity can also be
performed on obtaining $\ket{0}_b$ in the mapping back stage
followed by the correction of a known phase factor. For the
outcomes $\ket{\pm}\ket{-4\beta}_a$ and $\ket{\pm}\ket{4\beta}_a$
in the Bell state measurement, the CPB is prepared in states
$\ket{0}$ and $\ket{1}$ respectively, while for $\ket{-4\beta}_b$
and $\ket{4\beta}_b$ in the mapping back stage, it is prepared in
states $\ket{0}$ and $\ket{1}$ respectively. This completes our
teleportation protocol. The fidelity of the procedure is thus
unity with probability $1/4$, $\cos^2{\theta/2}$ with probability
$(3/8)\cos^2{\theta/2}$ and $\sin^2{\theta/2}$ with probability
$(3/8)\sin^2{\theta/2}$. Averaging over all possible initial
states one then gets an average fidelity of $3/4$, which is
greater than the classical teleportation fidelity of $2/3$.

    Let us clarify the sense in which the above is a bonafide
    teleportation procedure despite the systems being adjacent and
    the same CPB being {\em reused}. The CPB interacts with only cantilever
    $a$ during the Bell state measurement procedure and hence this
    can be considered as a local action by a party holding
    cantilever $a$. The CPB is automatically reset in the process as a fresh
    qubit not bearing any memory of its initial state. In the
    mapping back stage it can thus be regarded as a local device
    used by the party holding cantilever $b$ for extraction of the
    state.

Decoherence of the cantilever, if significant, will of course
affect both the generation of the state $|\psi(+)\rangle_{ab}$, as
well as the teleportation. However, decoherence of a cantilever is
in the coherent state basis and it will simply multiply the off
diagonal term $\ket{-2\beta}_a\ket{-2\beta}_b
\bra{2\beta}_a\bra{2\beta}_b$ (and its conjugate) in
$|\psi(+)\rangle_{ab}$ by a factor of the form $e^{-\Gamma}$ where
$e^{-\Gamma}\sim e^{-8\beta^2 \pi/Q}$ in which $Q$ is the quality
factor of the cantilevers \cite{blencowe} (note that as physically
expected, higher the quality factor, lower the decoherence).
Similarly, in evolutions given by Eq.(\ref{ev1}) and
Eq.(\ref{ev2}), the off diagonal terms
$\ket{0}\ket{0}_a\ket{2\beta}_b \bra{1}\bra{0}_a\bra{-2\beta}_b$
and $\ket{0}\ket{0}_b\bra{1}\bra{0}_b$ (and their conjugates) are
multiplied by $e^{-5\Gamma/2}$ and $e^{-\Gamma/2}$ respectively.
The net effect of decoherence at the end of the teleportation will
then be a reduction of fidelity corresponding to the
$\ket{\pm}\ket{0}_a$ outcome of the Bell state measurement to
$(2+e^{-4\Gamma})/3$, while the fidelity corresponding to other
outcomes will remain unchanged. Thus unless {\em all} coherence is
destroyed by decoherence {\em i.e.,} $e^{-4\Gamma}\sim 0$, we have
an average teleportation fidelity $2/3+e^{-4\Gamma}/12$, which is
better than $2/3$. For example, for $Q\sim 1000$ \cite{blencowe},
we have $e^{-\Gamma}\sim 0.975$ (for $\beta\sim 1$
\cite{blencowe}) and average teleportation fidelity is $0.74$. In
this paper we assume that the CPB hardly decoheres over the $ns$
time-scale of experiments with a GHz NC \cite{blencowe}.

\begin{figure}
\begin{center}
\includegraphics[width=3.5in, clip]{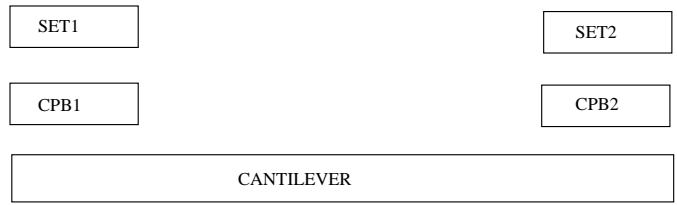}
 \caption{The figure shows a schematic diagram of the setup for entangling two Cooper pair boxes, denoted as CPB1 and CPB2
 respectively, through a cantilever.
 For the entangling procedure, no measurements are required.
 For verification of the entanglement through a witness, measurements need to be performed on CPB1 and CPB2 through the single electron
 transistors SET1 and SET2 respectively.}
\label{cpbent}
\end{center}
\end{figure}

 {\em Entangling two CPBs:}  The setting of our scheme of entangling two CPBs as
   depicted in Fig.\ref{cpbent} is two CPBs coupled
   to a single NC. The Hamiltonian for this system, in the absence of the voltage pulse
   giving rise to $-\frac{E_J}{2}\sigma_x$,  is
   well approximated (by
   straightforward extrapolation of Ref.\cite{blencowe}) as
   \begin{equation}
H=-2E_C(\sigma_z^{(1)}+\sigma_z^{(2)})+\hbar\omega_m a^{\dagger} a
+\lambda (a+a^{\dagger})(\sigma_z^{(1)}+\sigma_z^{(2)})
\end{equation}
$\sigma_z^{(i)}$ is a Pauli-Z operator of the $i$th Cooper pair
box, $a,a^{\dagger}$ are the annihilation-creation operators of
the nano-cantilever. We initially consider the NC to be starting
in the coherent state $|\alpha\rangle$ (we shall generalize later
to a thermal state) and the CPB's to be initialized in the state
$\ket{0}_1\ket{0}_2$, where labels $1$ and $2$ stand for the two
CPBs. At first, the Hamiltonian $-\frac{E_J}{2}\sigma_x$ is
applied to each CPB to rotate their states from $\ket{0}$ to
$\frac{1}{\sqrt{2}}(\ket{0}+\ket{1})$. Then evolution according to
the Hamiltonian $H$ kicks in and in a time $T=\pi/\omega_m$ the
evolution of the state can be calculated from
Ref.\cite{blencowe-review} to be
\begin{eqnarray}
\frac{1}{\sqrt{2}}(\ket{0}_1+\ket{1}_1)\frac{1}{\sqrt{2}}(\ket{0}_2+\ket{1}_2)|\alpha\rangle\rightarrow
\nonumber \\
\frac{1}{2}\{e^{-i(E_C
T+\phi(T,\beta,\alpha))}\ket{0}_1\ket{0}_2\ket{-\alpha-4\beta}\nonumber\\
+(\ket{0}_1\ket{1}_2+\ket{1}_1\ket{0}_2)\ket{-\alpha}\nonumber\\
+e^{i(E_C
T-\phi(T,\beta,\alpha))}\ket{1}_1\ket{1}_2\ket{-\alpha+4\beta}\},
\label{cpb}
\end{eqnarray}
where $\phi(T,\beta,\alpha)=2\beta\mbox{Im}\alpha$ is a phase
factor and $\ket{-\alpha}$,$\ket{-\alpha-4\beta}$ and
$\ket{-\alpha+4\beta}$ are coherent states. The sign flip from
$\alpha \rightarrow -\alpha$ in the above evolution occurs due to
the oscillator evolution for half a time period. The production of
states of the above type has been noted earlier in the context of
cavity-QED \cite{solano} and very recently in the context of
measurement based quantum computation \cite{spiller}. In
Ref.\cite{spiller}, it has been pointed out that for a large
$\beta$, a measurement of the oscillator (NC in our case) will
project the two qubits (CPBs in our case) probabilistically to the
maximally entangled state
$\ket{\psi^{+}}_{12}=\frac{1}{\sqrt{2}}(\ket{0}_1\ket{1}_2+\ket{1}_1\ket{0}_2)$.
Such an entangled state can, of course, be verified through Bell's
inequalities by measurements on the CPBs. However, here we want to
go beyond this result and reduce the requirements necessary for
observing entanglement between the CPBs. Suppose the cantilever
 is in a high temperature
thermal state so that position measurements of the cantilever
would be inefficient due to thermal noise. We thus ask the
question as to whether we can observe any entanglement between the
CPBs without the extra complexity of measurements on the NC. The
reduced density matrix of the two CPBs, when the states of the NC
are traced out will, for $\beta$ large, be
\begin{equation}
\rho_{12}=\frac{1}{4}(\ket{00}\bra{00}_{12}+\ket{11}\bra{11}_{12})
+\frac{1}{2} \ket{\psi^{+}}\bra{\psi^{+}}_{12}.
\label{cpbro}
\end{equation}
In deriving the above we have taken the overlap of coherent states
$\ov{-\alpha|-\alpha-4\beta}$,$\ov{-\alpha|-\alpha+4\beta}$ and
$\ov{-\alpha+4\beta|-\alpha-4\beta}$ to be nearly zero. Note that
when decoherence of the states of the cantilever is taken into
account, as it occurs in the coherent state basis \cite{blencowe},
we can, without loss of generality, replace
$\ket{-\alpha}$,$\ket{-\alpha-4\beta}$ and $\ket{-\alpha+4\beta}$
in Eq.(\ref{cpb}) by
$\ket{-\alpha}\ket{\xi_{-\alpha}}$,$\ket{-\alpha-4\beta}\ket{\xi_{-\alpha-4\beta}}$
and $\ket{-\alpha+4\beta}\ket{\xi_{-\alpha+4\beta}}$ where
$\ket{\xi_{-\alpha}}$,$\ket{\xi_{-\alpha-4\beta}}$ and
$\ket{\xi_{-\alpha+4\beta}}$ are three distinct environmental
states with pair-wise mutual overlap tending to zero in the limit
of strong decoherence. Thereby, for $\beta\sim 1$, the reduced
density matrix of the two CPBs is {\em unaffected} by decoherence
and still given by $\rho_{12}$ of Eq.(\ref{cpbro}). Also, note
that $\rho_{12}$ does not, in any way, depend on the initial
coherent state amplitude $\alpha$. Thus even if we were to start
in a thermal state of the cantilever given by $\int d^2\alpha~
P(\alpha)|\alpha\rangle\langle\alpha|$, the state of the two CPB
qubits for large $\beta$ will be $\rho_{12}$ for a time
$T=\pi/\omega_m$.

{\em Verification of the entanglement of the CPBs:}   $\rho_{12}$
is an entangled state, but not one that violates a Bell's
inequality. So we have to check the entanglement of the CPBs
through an entanglement witness
  \cite{sanpera}. Basically one has to measure the expectation value of the
  operator \cite{sanpera}
  \begin{equation}
  W=\frac{1}{4}\{I\otimes I+\sigma_z\otimes\sigma_z-\sigma_x\otimes\sigma_x-\sigma_y\otimes
  \sigma_y\}
  \label{w}
  \end{equation}
 for the state of the CPB qubits. The expectation value of $W$ is positive for
 all separable states, so if it is found to be negative, then we can conclude
 that the CPBs are entangled. In fact, for the predicted state $\rho_{12}$ at
 $T=\pi/\omega_m$, the
 expectation value of $W$ is $-0.25$. Note that the operator $W$ is a locally
 decomposable witness \cite{sanpera} which means that it is measurable
 by measuring only local operators in the same manner as Bell's inequalities.
 Its locally decomposable form is evident from Eq.(\ref{w}). Thus no
 interactions between the CPBs are needed to verify their entanglement, and they
 can well be beyond the range of each other's interactions. We have thus
 proposed a way of entangling two CPBs through a
 cantilever in thermal state in the presence of decoherence without using
 any measurements. This is an useful alternative to entangling the CPBs by direct
 interaction, as it will work even when the CPBs fall outside the range of each
 other's interaction. We have also proposed a method to verify their
 entanglement through local measurements on each of the CPBs. Of
 course, if the CPBs were allowed to resonantly exchange energy
 with a mode of the cantilever in analogy with Ref.\cite{cleland},
 then not only entanglement, but any quantum computation would be
 possible in of low decoherence \cite{cleland}. The presence of
 arbitrarily strong decoherence will, however,
 affect such a method. What we have shown is that even given the Hamiltonian of
 Ref.\cite{blencowe}, arbitrarily strong decoherence and thermal states,
 entanglement between the CPBs is still possible.

{\em Conclusions:} In this paper we have proposed a scheme to
entangle two mesoscopic systems of the same type through a third
mesoscopic system. In this context we have also proposed a
teleportation experiment in the mesoscopic setting using
continuous variable entanglement for discrete variable
teleportation.

   SB gratefully acknowledges a visit to Oklahoma State University during which this
   work started.

%______________________________________________________________________ BIBLIOGRAPHY

\end{document}